\documentclass[graphicx]{elsart}

\usepackage{graphicx}
\begin{document}
\begin{frontmatter}
\title{On the coefficients of the liquid drop model mass formulae and nuclear radii}

\author{G. Royer}\footnote{E-mail:royer@subatech.in2p3.fr} 
\address{Laboratoire Subatech, UMR: IN2P3/CNRS-Universit\'e-Ecole des Mines, \\
 4 rue A. Kastler, 44307 Nantes Cedex 03, France}

\begin{abstract}
The coefficients of different mass formulae derived from the liquid drop model and including or not the curvature energy, the diffuseness correction to the Coulomb energy, the charge exchange correction term, different forms of the Wigner term and different powers of the relative neutron excess $I=(N-Z)/A$ have been determined by a least square fitting procedure to 2027 experimental atomic masses. 
The Coulomb diffuseness correction $Z^2/A$ term or the charge exchange correction $Z^{4/3}/A^{1/3}$ term plays the main role to improve the accuracy of the mass formula. The Wigner term and the curvature energy can also be used separately for the same purpose. The introduction of an $|I|$ dependence in the surface and volume energies improves slightly the efficiency of the expansion and is more effective than an $I^4$ dependence. Different expressions reproducing the experimental nuclear charge radius are provided. The different fits lead to a surface energy coefficient of around 17-18~MeV and a relative equivalent rms charge radius r$_0$ of 1.22-1.23~fm.

\noindent {\small {\em PACS:} 21.10.Dr; 21.60.Ev; 21.60.Cs }

\end{abstract}
\end{frontmatter}

\newpage

\section{Introduction}
The binding energy of new nuclides both in the superheavy element region and the regions close to the 
proton and neutron drip lines are still poorly known and the different theoretical extrapolations do not fully agree. Therefore other studies are still necessary to better predict the masses of such exotic nuclei. The atomic nucleus resembling to a charged liquid drop, 
semi-macroscopic models including a pairing energy term have been firstly advanced to determine the nuclear masses 
\cite{wei35,bet36}.  
Macroscopic-microscopic approaches, mainly the finite-range droplet model and the finite-range liquid drop model \cite{moll95} have been proposed to simulate the smooth part of the nuclear masses and the non smooth part depending on the parity of the proton and neutron numbers and the proximity of the magic numbers. Nuclear masses have also been reproduced accurately within the statistical Thomas-Fermi model within a Seyler-Blanchard effective interaction \cite{ms94rep,ms96}. Microscopic 
Hartree-Fock self-consistent calculations using the mean-field approach and Gogny or Skyrme forces and pairing correlations  
 \cite{samy02,sto05} as well as relativistic mean field theories \cite{bend01} have also been advanced to try to reproduce 
 these nuclear masses. Neural networks have been recently used \cite{ath04} in the same purpose.

Beyond the description of the nuclear ground state energy, the evolution of the nuclear binding energy with deformation and rotation governs the fission, fusion, cluster 
and $\alpha$ decay potential barriers and the existence of the well deformed rotating states.
Within the liquid drop model approach, the main characteristics of these barriers may be reproduced using,  
firstly, four basic macroscopic terms : the volume, surface, Coulomb and nuclear proximity energy terms and, secondly,
shell and pairing energy contributions to explain structure effects and 
improve quantitatively the results \cite{rr84,rr85,roy00,rm01,rg03,rbon06}.

In a first paper \cite{roygau} some coefficients and terms of the liquid drop model mass formula and a specific nuclear radius have been fitted using a set of 1522 experimental nuclear masses. The purpose of the present work is to extend this investigation to determine the relative efficiency of different more complex combinations of terms 
of the liquid drop model to reproduce the  
masses \cite{aud03} of 2027 nuclei and to study, particularly, the separated influence of the curvature energy, the different forms of the Wigner term and of different powers of the relative neutron excess $I=(N-Z)/A$. 
As suggested in \cite{lunn03} fits on ground state masses alone have been prefered as was done with the Hartree-Fock mass formulae to avoid that the parameters might be distorted by the theoretical and experimental uncertainties associated with the barriers. The second aim is to fit the experimental nuclear charge radii by different expressions and to compare with the relative charge radii $r_0=R_0/A^{1/3}$ derived from the different mass formulae. Finally, the coefficients of the mass formulae using these different expressions for the charge radius are determined as well as their accuracy. Another motivation of this work is to improve later the coefficients of the generalized liquid drop model previously proposed \cite{rr85}. Recently, the mutual influence of terms in semi-empirical formulae has been deeply investigated \cite{kirs08}. 

\section{Nuclear binding energy}
The nuclear binding energy $B_{nucl}$(A,Z) is the energy needed for separating all 
the nucleons constituting a nucleus. It
is related to the nuclear mass $M_{n.m}$ by 
\begin{eqnarray}                 
B_{nucl}(A,Z)=Zm_P+Nm_N-M_{n.m}(A,Z).
\end{eqnarray}
B$_{nucl}$(A,Z) may thus be connected to the experimental atomic masses given in \cite{aud03} since :
\begin{eqnarray}                 
M_{n.m}(A,Z)=M_{a.m}(A,Z)-Zm_e+B_e(Z).
\end{eqnarray}
The binding energy B$_e$(Z) of all removed electrons is \cite{lunn03}  
\begin{eqnarray}              
B_e(Z)=a_{el}Z^{2.39}+b_{el}Z^{5.35},
\end{eqnarray}
with $a_{el}=1.44381\times10^{-5}$~MeV and 
$b_{el}=1.55468\times10^{-12}$~MeV.
 
Different subsets of the following expansion of the nuclear binding energy in powers of $A^{-1/3}$ and $\vert I \vert$ have been considered :
\begin{eqnarray}                 
B = a_v \left(1-k_{v_1}\vert I \vert-k_{v_2}I^2-k_{v_3}I^4\right)A
-a_s\left(1-k_{s_1}\vert I \vert-k_{s_2}I^2-k_{s_3}I^4\right)A^{\frac {2}{3}} \nonumber \\ 
-a_k\left(1-k_{k_1}\vert I \vert-k_{k_2}I^2-k_{k_3}I^4\right)A^{\frac {1}{3}}   
-a_0A^0-\frac {3}{5} \frac {e^2Z^2}{R_0}+f_p \frac {Z^2}{A}+a_{c,exc} \frac {Z^{\frac {4}{3}}}{A^{\frac {1}{3}}}~ \nonumber \\
-E_{pair}-E_{shell}-E_{Wigner}.~~~~~~~~~~~~~~~~~~~~~~~~~~~~~~~~~~~~~~~~~~~~~~~~~~~~~~~ 
\end{eqnarray}  
The first term gives the volume energy corresponding to the saturated exchange force and infinite nuclear matter.
$I^2A$ is the asymmetry energy of the Bethe-Weizs\"acker mass formula.
 The second term is the surface energy. It takes into account the deficit of binding energy 
of the nucleons at the nuclear surface and corresponds to semi-infinite nuclear matter. The Bethe-Weizs\"acker 
mass formula does not consider the dependence of the surface energy on $I$. This term was originally contained in the Weizs\"acker formula \cite{wei35}. The third term is the curvature energy.
It is a correction to the surface energy resulting from local properties 
and consequently depending on the mean local curvature. This term is considered in the 
Lublin-Strasbourg Drop (LSD) model \cite{lsdm}, the TF model \cite{ms96} but not in the FRLDM \cite{moll95}.
In the three first  terms a dependence on $\vert I \vert$ and $I^4$, the so-called malacodermous term, has been envisaged
since they have been proposed to better reproduce the fission barrier heights \cite{dahl82} and to simulate the softening
of the surface of highly neutron-rich nuclei \cite{dutt86}. 
The A$^0$ term appears when the surface term is extended to include higher order terms in 
A$^{-1/3}$ and $I$.  
The fifth term gives the decrease of binding energy due to the repulsion between the protons.
 In the Bethe-Weizs\"acker mass 
formula this Coulomb energy is proportional to $Z(Z-1)$. Different formulae will be assumed for the charge radius.
 The $Z^2/A$ term is the diffuseness correction to the basic sharp radius Coulomb energy term (called also the proton form-factor correction to the Coulomb energy in \cite{moll95}) and the term proportional to $Z^{4/3}/A^{1/3}$ is the charge exchange correction term. 

The pairing energy has been calculated with the following expressions used for spherical nuclei in the recent version of the Thomas-Fermi model \cite{ms96}.\\
For odd Z, odd N and N=Z nuclei
\begin{eqnarray}  
E_{Pair}=4.8/N^{1/3}+4.8/Z^{1/3}-6.6/A^{2/3}-30/A~MeV.
\end{eqnarray}
For odd Z, odd N and $N \ne Z$ nuclei
\begin{eqnarray}  
E_{Pair}=4.8/N^{1/3}+4.8/Z^{1/3}-6.6/A^{2/3}~MeV.
\end{eqnarray}
For odd Z, even N nuclei
\begin{eqnarray}  
E_{Pair}=4.8/Z^{1/3}~MeV.
\end{eqnarray}
For even Z, odd N nuclei
\begin{eqnarray}  
E_{Pair}=4.8/N^{1/3}~MeV.
\end{eqnarray}
For even Z, even N nuclei
\begin{eqnarray}  
E_{Pair}=0.
\end{eqnarray}

The theoretical shell effects used in the TF model ($7^{th}$ column of the table in \cite{ms94rep} and \cite{ms96}) 
have also been retained since they 
allow to reproduce correctly the masses from fermium to $Z=112$ \cite{hof96}. They have been calculated from the Strutinsky 
shell-correction method and previously to the other coefficients of the TF model. The fits on nuclear masses depend necessarily on the choice of the selected theoretical shell effects or the formulae chosen to describe these shell effects. The sign for the shell energy term comes from the adopted definition in \cite{ms94rep}. 
It gives, for example, a contribution of $12.84~$MeV to the binding energy of $^{208}$Pb. 

The Wigner energy allows to reproduce the kink in the nuclear mass surface that is not a shell effect in the usual sense. It depends on $I$ and appears in the counting of identical pairs in a nucleus. Different expressions are considered. The first  expression is simply $W\vert I \vert$ \cite{mye77}. Its effect is to decrease the binding energy when $N \ne Z$. 

The congruence energy term is given by :
\begin{eqnarray}                 
E_{cong}=-10exp(-4.2\vert I \vert)~MeV.
\end{eqnarray} 
It represents an extra binding energy associated with the presence of congruent pairs \cite{ms96}. 

Within an Hartree-Fock approach \cite{gori02} it has been assumed that there is nothing compelling about an exponential representation and a gaussian expression 
\begin{eqnarray}                  
E=V_W~exp(-\lambda I^2)
\end{eqnarray}                
is just as acceptable.

Another term has also been proposed in \cite{gori02} 
\begin{eqnarray} 
E=\beta \vert N-Z \vert~exp\left\lbrack -(A/A_0)^2 \right \rbrack.
\end{eqnarray}
We have also tested separately two other possible expressions
\begin{eqnarray}                  
E=\beta \vert N-Z \vert~exp\left\lbrack -(A/A_0) \right \rbrack.     
\end{eqnarray}
and 
\begin{eqnarray}                 
E=\beta \frac{\vert N-Z \vert}{1+ (A/A_0)^2}.
\end{eqnarray}
The term of nuclear proximity energy does not appear in the binding energy of the ground state since it becomes effective  
only for necked shapes but not for slightly deformed ground states.
 
\section{Coefficients of the mass formulae}
 To obtain the coefficients of the different expansions by a least square fitting procedure, the masses of the 2027 nuclei verifying the two conditions : N and Z higher than 7 and the one
 standard deviation uncertainty on the mass lower than or equal to 150 keV \cite{aud03} have been used. These restrictions
 are not employed by all investigators \cite{kirs08,diep07}.
The root-mean-square deviation $\sigma$ defined by 
\begin{eqnarray} 
\sigma ^2= \frac {\Sigma \left \lbrack  M_{Th}- M_{Exp}\right 
\rbrack ^2}{n}
\end{eqnarray}
has been used to determine the relative efficiency of the different selected sets of terms since $n\gg f$ where f is the number of fit parameters. A very efficient software has been used. The extraction of standard errors on the fit parameters seems very difficult but the errors are surely very weak. On the other hand, the values of the last decimals of a coefficient can be changed in counterbalancing by a change in the last decimals of another coefficient for almost the same rms deviation.   

In  Table~I, the improvement of the nuclear mass reproduction when additional contributions
are added to the basic $A,~AI^2,~A^{2/3},~A^{2/3}I^2,~Z^2/A^{1/3}$ terms is clearly displayed. The curvature energy is not taken into account. The introduction of the pairing term is obviously necessary. The introduction of a constant term improves slightly the adjustment and changes strongly the surface energy coefficient. It induces also a severe discontinuity during the transition from one to two-body shapes as in fission, fusion or $\alpha$ emission.  The congruence energy term at least with the fixed coefficients adopted here (as in the LSD and TF models) is much less efficient to lower $\sigma$ than the Wigner term W$\vert I \vert$. When the coefficients before the exponential and the exponent
are free the congruence energy tends to the usual Wigner term since the coefficient before the exponential diminishes 
while the exponent increases. The diffuseness correction to the Coulomb energy and the standard Wigner term W$\vert I \vert$ can be used separately to strongly lower $\sigma$.  
The Wigner energy is approximately independent of the nuclear shape  \cite{mye77}. In a division into 2 fragments, all with the same value of $\vert I \vert$, the form W$\vert I \vert$ for the Wigner energy jumps at scission
to 2 times its original value, which leads to a discontinuity of around 6.7 MeV of the potential energy at the scission point between the nascent fragments for $^{258}$Fm as an example. For the congruence energy the discontinuity is also important : 3.9 MeV for this same nucleus. The Coulomb diffuseness correction term has the main advantage to be almost continuous at the scission point in the entrance or exit channels. The combination of the Coulomb diffuseness correction term and the W$\vert I \vert$ term allows to reach the very satisfactory value $\sigma=0.608$ \cite{moll95,sto05,lsdm}.  
 The $A^{2/3}|I|$ term is useful to improve the accuracy of the expansion and is more effective than the $A^{2/3}I^4$
 term but when the Wigner term and the Coulomb diffuseness correction factor term are taken into account
the introduction of the $A^{2/3}|I|,~A^{2/3}I^4$ and $A^0$ terms are ineffective. 

In  Table~II, the efficiency of the curvature
 energy term with different $I$ dependences is investigated without taking into account the Wigner contribution. 
When the Coulomb diffuseness correction is disregarded the introduction of the term in $A^{1/3}$ allows to decrease $\sigma$ of 0.1 MeV. When the Coulomb diffuseness correction is considered the term in $A^{1/3}$ is ineffective
. The addition of  a $A^{1/3}I^2$ term 
 improves slightly the results. Supplementary terms in $|I|$ to calculate the volume, surface and curvature
energies allow finally to reach $\sigma$=0.58~MeV. They are still more efficient than the $I^4$ terms.
The curvature energy has the advantage to be continuous at the scission point at least 
in symmetric entrance or decay channels. It has the disadvantage that its value (and its sign) lacks of stability.   

These two first tables show a good convergency of the volume a$_v$ and asymmetry volume k$_v$ constants 
respectively towards around 15.5~MeV and $1.7-1.9$ . The variation of the surface coefficient is larger
 but a$_s$ evolves around 17-18~MeV. Small values of the surface coefficient favors quasi-molecular or 
two-body shapes at the saddle-point of the potential barriers while large values of $a_s$ promote 
one-body elongated shapes. As it is well known the surface asymmetry coefficient k$_s$ is less easy to precise.

For the Bethe-Weizs\"acker formula the fitting procedure leads to  
\begin{eqnarray}                 
B_{nucl}(A,Z)=15.5704A-17.1215A^{2/3}\\
-0.71056\frac {Z(Z-1)}{A^{1/3}}-23.4496I^2A 
-E_{pair}-E_{shell} \hfill \nonumber
\end{eqnarray}
with $\sigma$=1.30~MeV. That gives r$_0$=1.216~fm and k$_v$=1.506. The $\sigma$ value is explained by the non dependence of the surface energy term on the relative neutron excess $I$.

The mass formulae including a Wigner term given by the formulae (12),(13) or (14) are examined in Table~III.
The values of $A_0$ which minimize the mass rms deviation are respectively 48, 35 and 40 for these three expressions. 
They are determined with an accuracy of around 5 mass numbers.
These formulas for the Wigner energy are supposed to be approximately independent of the nuclear shape. Their discontinuity at the scission point of fission or fusion barriers is less important than that of the congruence and 
 W$\vert I \vert$ terms. For example, for the expressions 
$\vert N-Z \vert~exp\left\lbrack -(A/48)^2 \right \rbrack$, $1.5\vert N-Z \vert~exp\left\lbrack -(A/35) \right \rbrack$   
and $1.6\frac{\vert N-Z \vert}{1+ (A/40)^2}$ and symmetric decay of the $^{254}$Fm the discontinuities at the contact point of the nascent fragments are respectively 0, 0.05 and 2.2 MeV.

The first, fifth and eighth lines of the table indicate that within the simplest form for the volume and surface energies
, thus with only 7 parameters, the r.m.s deviation on the masses is less than 0.6 MeV. The introduction of a dependence on  $\vert I \vert$ and of the curvature energy lowers $\sigma$ of 0.04~MeV within the expression (12) and of 0.02~MeV for the two last ones. The surface energy coefficients are generally lower with these forms of the Wigner energy and, consequently,  $r_0$ is generally higher.

Within the Wigner term derived from the Hartree-Fock approach \cite{samy02} and taking $A_0=28$ and $\lambda=485$ in the formulas (11) and (12) as derived using the BSk2 Skyrme force the following formula is obtained
\begin{eqnarray}                 
B=15.4503 \left(1-1.7463I^2\right) A
-17.5701\left(1-1.5296I^2\right) A^{2/3}  \nonumber \\
-\frac {0.6e^2Z^2}{1.219A^{1/3}}+1.1948\frac {Z^2}{A}
 -E_{pair}-E_{shell}+1.769 e^{-485 I^2} \nonumber \\
-0.2197\vert N-Z \vert e^{-(A/28)^2}~~~~~~~~~~~~~~~~~~~~~~
\end{eqnarray}
with $\sigma=0.623~MeV$.

The accuracy can be improved in varying $A_0$ and $\lambda$ : 
\begin{eqnarray}                 
B=15.4121 \left(1-1.7972I^2\right) A  
-17.3059\left(1-1.7911I^2\right) A^{2/3}  \nonumber \\
-\frac {0.6e^2Z^2}{1.232A^{1/3}}+0.8961\frac {Z^2}{A} 
-E_{pair}-E_{shell}+2.25 e^{-80 I^2} \nonumber \\
-0.4883\vert N-Z \vert e^{-(A/50)^2}~~~~~~~~~~~~~~~~~~~~.
\end{eqnarray}
It leads to $\sigma=0.573~MeV$ and produces sizable changes in several of the other parameters which proves the importance of the Wigner term and the mutual influence of terms in the semi-empirical mass formulae \cite{kirs08}.

In Table IV the efficiency of the charge exchange correction term in $Z^{4/3}/A^{1/3}$ is studied. The lines can be compared respectively to the $7^{th}$ line of table 1, $2^{nd}$ line of table 2, $8^{th}$ line of table 1, $3^{rd}$ line of table 2, $6^{th}$ and $12^{th}$ lines of table 1 and $1^{st}$ and $5^{th}$ lines of table 3. The only change is that the diffuseness correction term in $Z^2/A$ has been replaced by the charge exchange correction term. It is quite interesting to observe that  
the introduction of these two terms separately leads to the same accuracy, the same value of the charge radius, almost the same values of the surface coefficients and to small changes of the volume coefficients. So the charge exchange correction term is as efficient as the diffuseness correction term and there is no correlation between these two terms and the charge radius and the surface coefficients and weak correlation with the volume coefficients. The introduction of both these two terms does not allow to improve the accuracy of the mass formulae and leads to spurious values of the volume coefficient. 

Finally, the shell and pairing energies, the diffuseness correction term to the Coulomb energy, the charge exchange correction term and the Wigner energy separately lead to significant reductions of the rms deviation while the other   
terms increase the accuracy by only 2 or 3 per cent. A correlation between the surface coefficient and the radius can be extracted. The relative radius $r_0$ diminishes when the surface coefficient increases.

\section{Nuclear charge radius}
Experimentally and for nuclei verifying N and Z higher than 7 the 
set of 782 ground state nuclear charge radii presented in ref. \cite{ange04} indicates a rms charge radii of $0.94944A^{1/3}$ which leads for the equivalent rms charge radius (denoted by Q in ref. \cite{hamy88}) given by
\begin{eqnarray}                 
R_0=\sqrt{\frac{5}{3}}<r^2>^{1/2}
\end{eqnarray}
to the value  $R_0=1.2257~A^{1/3}~fm$ with $\sigma =0.124~fm$.
Other data have been obtained recently \cite{libe07}.
In the adjustment to the nuclear masses displayed in the tables I and II  the reduced charge radius r$_{0, charge}$ converges to 1.22-1.23~fm in good agreement with the experimental data for the charge radius. The introduction of the expressions (12), (13) and (14) for the Wigner term leads to slightly higher values.

The experimental data indicate that the ratio $R_0/A^{1/3}$ is not strictly constant. For example,  
$R_0/A^{1/3}=1.312~fm$ for $^{40}$Ca and $R_0/A^{1/3}=1.234$ for $^{48}$Ca
 while $R_0/A^{1/3}=1.217$  for $^{190}$Pb and $R_0/A^{1/3}=1.201~fm$ for $^{214}$Pb.
These $R_0$ and $r_0$ values correspond to the Coulomb energy of a charged liquid drop with constant charge density and spherical sharp surface. It must not be confused with half-density radius derived from two-parameter Fermi function or other parameters entering in other parametrisations of the charge distribution. 

In the adjustment to the experimental nuclear masses the nuclear mass radius is not fitted. Root-mean-squared matter radii are given in Ref. \cite{lima04} for specific nuclei. For all isotopic series a decrease of the mass rms radius is observed with increasing neutron number as for the charge radius. In this mass range ($A=63-75$) the radius of the neutron and proton distributions are very similar. Global fits lead to an overall contraction of the nuclear radius as $T=\vert N-Z \vert/2$ increases \cite{dufl94}, a so called "isospin shrinkage". Recently, an estimate of the neutron skin linear in the relative neutron excess I has been extracted from experimental proton radii and observed mirror displacement energies \cite{dufl02} and the coefficients have been also determined from antiprotonic X-rays and radiochemical data \cite{frie05}.

Within a simple form of the mass formula including the Coulomb diffuseness correction and the Wigner term given by the formula (13) but not the dependence on the curvature energy and on $\vert I \vert$ and $I^4$ the introduction of the factor $r_0=1.2257$ deduced from the experimental charge radius leads to 
\begin{eqnarray}                 
B=15.3543 \left(1-1.7445I^2\right) A 
-17.2293\left(1-1.5765I^2\right) A^{2/3}  \nonumber \\
-\frac {0.6e^2Z^2}{1.2257A^{1/3}}+1.2442\frac {Z^2}{A} 
-E_{pair}-E_{shell}-0.7641\vert N-Z \vert e^{-A/35}
\end{eqnarray}
and $\sigma =0.615~MeV$.

More accurate formulae may be used to reproduce the mean behaviour of the equivalent rms charge radius.

The expression
\begin{eqnarray}                  
R_0=1.0996A^{1/3}+0.653~fm
\end{eqnarray}
leads to $\sigma =0.066~fm$. Its introduction in the above-mentioned mass formula curiously diminishes the accuracy 
to $\sigma =1.16~MeV$. The same behaviour is observed when the dependence in $Z^2/A$ is replaced by a dependence in $Z^{4/3}/A^{1/3}$.

The formula  
\begin{eqnarray}                 
R_0=1.1718A^{1/3}+\frac {1.4069}{A^{1/3}}~fm
\end{eqnarray}
leads to $\sigma =0.064~fm$, while for the corresponding mass formula $\sigma =0.929~MeV$.

The adjustment to the experimental charge radii within the form proposed in Ref.\cite{bloc77} gives  
\begin{eqnarray}                
R_0=1.1818A^{1/3}-0.089+\frac {1.5938}{A^{1/3}}~fm
\end{eqnarray}
with $\sigma =0.064~fm$ and $\sigma =0.900~MeV$ while the form selected in Ref.\cite{seeg75} leads to 
\begin{eqnarray}                  
R_0=1.1769A^{1/3}+\frac{1.2046}{A^{1/3}}+\frac {1.5908}{A}~fm
\end{eqnarray}
and $\sigma =0.064~fm$ for the radius and $0.88~MeV$ for the corresponding mass formula.

The form of the expression given the equivalent rms charge radius proposed in Ref.\cite{myer00}
\begin{eqnarray}                  
R_0=1.2332A^{1/3}+\frac{2.8961}{A^{2/3}}-0.18688A^{1/3}I~fm
\end{eqnarray}
gives a good accuracy $\sigma =0.052~fm$ but $\sigma =0.765~MeV$.

The fact that improved formulae for the charge radius lead to significantly poorer fits to the masses calls into question the reliability of the equivalent uniform charge distribution as a link between charge radii and Coulomb energy. 

Finally a rms deviation for the charge radius of only 0.01 fm (fitted on 362 nuclei) has been obtained in Ref. \cite{dufl94}
in expressing the charge mean square radius by 
\begin{eqnarray}                  
r^3=a+bA+cT
\end{eqnarray}
for spherical nuclei and taking into account the deformation for the other nuclei via a nucleonic promiscuity form factor P depending on the distance of Z and N to the proton and neutron magic numbers.    

To describe the main properties of the fusion, 
fission, cluster and $\alpha$ emission
potential barriers in the quasi-molecular shape path,  the formula
\begin{eqnarray} 
R_0=1.28A^{1/3}-0.76+0.8A^{-1/3}
\end{eqnarray}
 proposed in Ref.\cite{bloc77} for the equivalent rms radius has been retained
in the GLDM \cite{rr84,rr85,roy00,rm01,rg03,rbon06,roygau}. This formula coming from the Droplet Model approach
simulated a small decrease of the density with increasing mass. This does not seem corroborated by the 
experimental data \cite{ange04} at least on the charge radius.  
Nevertheless without changing the surface and Coulomb energies which are shape-dependent terms, the adjustment of the volume energy 
coefficients and the introduction of a pure Wigner term leads to
\begin{eqnarray} 
B=15.5209 \left(1-1.934I^2\right) A 
-17.9439\left(1-2.6I^2\right) A^{2/3}  \nonumber \\
-\frac {0.6e^2Z^2}{1.28A^{1/3}-0.76+0.8A^{-1/3}}+2.1798\frac {Z^2}{A} 
 -E_{pair}-E_{shell}-27.21 \vert I \vert~~~~~~
\end{eqnarray}
and $\sigma=0.686~MeV$.
The introduction of other terms can still allow to diminish $\sigma$.

The Fig. 1 displays the difference between the theoretical and experimental masses within a formula of Table I recalled below : 
\begin{eqnarray}                 
B=15.3543 \left(1+0.0284\vert I \vert-1.8837I^2\right) A   \nonumber \\
-17.0068\left(1+0.131\vert I \vert-2.274I^2\right) A^{2/3}  \nonumber \\
-\frac {0.6e^2Z^2}{1.228A^{1/3}}+1.0339\frac {Z^2}{A} 
 -E_{pair}-E_{shell}-16.27 \vert I \vert~~~~~~~~~~
\end{eqnarray}
with $\sigma=0.605~MeV$.
For most of the nuclei with A higher than 110 the difference between the theoretical and experimental masses
is less than 1 MeV.
\begin{figure}[htbp]
\begin{center}
\includegraphics[height=5.5cm]{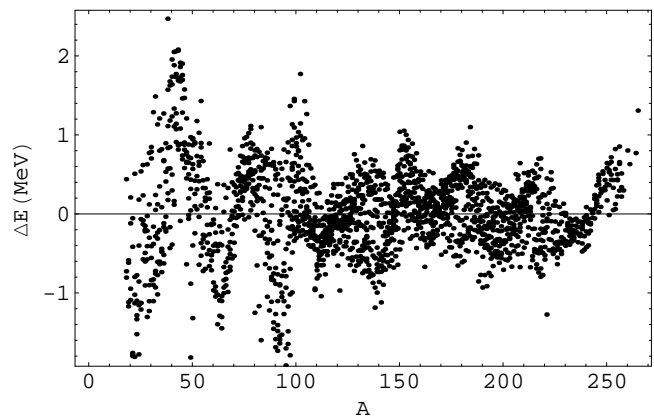}
\caption{Difference (in MeV) between the theoretical and experimental masses 
for the 2027 selected nuclei as a function of the mass number.}
\end{center}
\end{figure}

\section{Summary and conclusion}

The coefficients of different mass formulae derived from the liquid drop model and including or not the curvature energy, the Coulomb diffuseness correction energy, the charge exchange correction term, different forms of the Wigner term and different powers of the relative neutron excess $I$ have 
been determined by a least square fitting procedure to 2027 experimental atomic 
masses. 

The Coulomb diffuseness correction $Z^2/A$ term or the charge exchange correction $Z^{4/3}/A^{1/3}$ term plays the main role to improve the accuracy of the mass formula. The Wigner term and, with a smaller effect, the curvature energy can also be used separately for the same purpose but their coefficients are unstable. The introduction of an $|I|$ dependence in the surface and volume energies improves only slightly the efficiency of the expansion. An $I^4$ dependence and a constant term are not worth including. 
Expressions for the Wigner term leading to small discontinuities at the scission point of the potential barriers are as efficient as the W$\vert I \vert $ term. Different expressions reproducing the experimental nuclear charge radius are provided. The different fits lead to a surface energy coefficient of around 17-18~MeV and a relative equivalent rms charge radius r$_0$ of 1.22-1.23~fm.

\begin{table*}
\rotatebox{90}{
\caption{\label{tab:table1}Dependence of the energy coefficient values (in MeV or fm) on the selected term set
 including or not the pairing and congruence terms and root mean square deviation. The theoretical shell 
energies are taken into account but not the curvature energy. The Coulomb energy is determined by
$a_c \frac {Z^2}{A^{1/3}}$ with $a_c=3e^2/5r_0$.}
}
\rotatebox{90}{
\scalebox{.9}{
\begin{tabular}[horizontal]{ccccccccccccccc}
\hline
$  a_v$& $k_{v_1}$&$k_{v_2}$ &$k_{v_3}$&   $a_s$&$k_{s_1}$&$k_{s_2}$&$k_{s_3}$& $a_0$  &$W$       & $Cong$  &$Pair$& $r_0$ & $f_p$   &$\sigma$  \\	
\hline
15.5959&-         &  1.7051 & -      & 17.1723  &-        &0.994  &-        &-       &-        &n         &n        & 1.227 &-       &1.32 \\  
\hline
15.4996&-         &1.7011 & -       & 16.7628&-        &0.990 &-        &-       &-        &n         &y        & 1.235 &-&0.977  \\
\hline
15.0775&-         &1.6799 & -       & 14.8548&-&1.005&-&7.686 &-        &n         &y        & 1.269 &-&0.917  \\
\hline
15.93&-         &  1.8454 & -       & 18.7399&-        &1.722  &-        &-       &-        &y         &y        & 1.204 &-&0.870  \\
\hline
15.3416&-    &  1.8220 & -  & 16.0795&-        &1.838  &-        &10.982       &-        &y         &y        & 1.251 &-&0.728  \\
\hline
15.8559&-         &  1.8549  & -       &19.3863 &-        &1.999  &-        &- &-        &y         &y        & 1.191 &1.1478&0.628  \\
\hline
15.4108&-         &  1.7119 & -      &17.5361 &-       &1.404&  -         &-       &-        &n        &y         &1.218 &1.3755&0.660\\
\hline
15.4721&-         &  1.7154 & -       & 17.8591 &-        &1.411&-        &-1.2       &-        &n         &y        & 1.212 &1.4358&0.659  \\
\hline
15.4436&-         &  1.5909& 2.147& 17.4801&-  &0.732  &10.96  &-  &-        &n         &y        & 1.215 &1.2286&0.640  \\
\hline
15.4367&-0.0551&1.9205&-& 17.2772&-0.296  &2.420  &- &-  &-        &n         &y        & 1.215 &1.1095&0.616  \\
\hline
15.3306&-&1.8991&-& 16.1189&-  &2.410&- &-  &41.82 &n         &y        & 1.254&-&0.729 \\
\hline
15.351&-&1.8097&-& 16.9773&-  &1.972&- &-  &21.60        &n         &y        & 1.233 &0.9621&0.608 \\
\hline
15.3543&-0.0284&  1.8837& -& 17.0068 &-0.131&2.274&-        &-       &16.27 &n         &y        & 1.228 &1.0339&0.605  \\
\hline
15.5494&-         &  1.8406& -& 17.9723&-&2.077&-&-4.130       &26.08 &n         &y        & 1.216 &1.0681&0.590  \\
\hline
15.3619&- &  1.7539&0.799& 17.0142&-&1.682&3.88&-       &19.57 &n         &y        & 1.230 &0.9879&0.606  \\
\hline
\end{tabular}
}}
\end{table*}

\begin{table*}
\rotatebox{90}{
\caption{\label{tab:table2}Dependence of the energy coefficient values (in MeV or fm) on the selected term set
 and root mean square deviation. The theoretical pairing and shell 
energies are included but not the Wigner energy. The Coulomb energy is determined by
$a_c \frac {Z^2}{A^{1/3}}$ with $a_c=3e^2/5r_0$.}
}
\rotatebox{90}{
\scalebox{0.9}{
\begin{tabular}{ccccccccccccccccc}
\hline
$  a_v$& $k_{v_1}$&$k_{v_2}$ &$k_{v_3}$&   $a_s$&$k_{s_1}$&$k_{s_2}$&$k_{s_3}$& $a_k$  &$k_{k_1}$& $k_{k_2}$&$k_{k_3}$& $r_0$& $f_p$   &$\sigma$  \\	
\hline
14.772&-&   1.6753 & - & 12.3434&-        &1.094 &-  &7.602 &-        &-  &-  & 1.285 &-&0.873 \\
\hline
15.4623&-         &  1.7140 & -       & 17.8861&-        &1.402  &-        &-0.564  &-        &-  &-        & 1.214 &1.4157&0.660  \\
\hline
15.444&-         &  1.8730 & -       & 18.0077&-        &2.783  &-        &-1.269  &-        &47.98 &-        & 1.221 &1.3721&0.639  \\
\hline
15.4401& -        &  2.3713 & -7.447 & 17.9965& -       &7.136  &-65.24 &-1.983&-        &134.19 &-1530.7& 1.230 &1.1196&0.594 \\
\hline
15.7016& 0.1622 &  1.351 &-        & 19.6025& 1.223 & -1.445 &    -     &-5.116 & 11.917 &-27.66    &   -      &1.220& 1.0166& 0.583 \\
\hline
\end{tabular}
}}
\end{table*}

\begin{table*}
\rotatebox{90}{
\caption{\label{tab:table3}Dependence of the energy coefficient values (in MeV or fm) on the selected term set
 and root mean square deviation. The theoretical pairing and shell 
energies are included. The Wigner energy determined by the expressions (12), (13) or (14) are taken into account. 
The $I^4$ term is disregarded.}
}
\rotatebox{90}{
\scalebox{.92}{
\begin{tabular}{ccccccccccccccc}
\hline
$  a_v$& $k_{v_1}$&$k_{v_2}$ &$a_s$&$k_{s_1}$&$k_{s_2}$& $a_k$  &$k_{k_1}$& $k_{k_2}$
&$\beta~(12)$   &$\beta~(13)$   &$\beta~(14)$ & $r_0$& $f_p$   &$\sigma$  \\	
\hline
15.2565&-& 1.7475& 16.8485&-&1.603 &-  &-&-&0.679  &-  &-  & 1.241 &1.1161&0.600 \\
\hline
15.311&-0.035& 1.8745&16.826&-0.197&2.238&-  &-&-&0.535&-  &-  & 1.235 &0.9642&0.580 \\
\hline
15.5431&-& 1.6662&18.8035&-&0.721&-3.343&-&-12.95&1.017&-  &-  & 1.220 &1.2826&0.580 \\
\hline
15.8011&0.0752& 1.5133&20.0578&0.466&-0.223&-5.754&4.068&-15.75&0.794&-  &-  & 1.213 &0.9836&0.556 \\
\hline
15.2374&-&1.7880&16.6394&-&1.803&-&-&-&-&1.514&-  & 1.245&0.9639&0.590 \\
\hline
15.2949&-0.0047&1.8077&16.7004&-0.057&1.958&-&-&-&-&1.319&-  & 1.241&0.8456&0.585 \\
\hline
15.3474&0.1082&1.3103&17.3787&0.800&-2.111&-1.789&16.015&-85.97&-&2.249&-  & 1.247&0.7978&0.571 \\
\hline
15.2738&-&1.8238&16.6658&-&2.010&-&-&-&-&-&1.262  & 1.242&0.8822&0.581 \\
\hline
15.315&-&1.7455&16.7943&-&1.150&-0.217&-&-190.59&-&-&1.581&1.238&0.8382&0.574\\
\hline
15.265&0.1212&1.2866&16.8666&0.873&-2.401&-1.041&22.288&-154.51&-&-&2.516&1.254&0.7380&0.563\\
\hline
\end{tabular}
}}
\end{table*}

\begin{table*}
\rotatebox{90}{
\caption{\label{tab:table4}Dependence of the energy coefficient values (in MeV or fm) on the selected term set
 and root mean square deviation. The theoretical pairing and shell 
energies are included as well as the charge exchange correction term. The Coulomb energy is determined by
$a_c \frac {Z^2}{A^{1/3}}$ with $a_c=3e^2/5r_0$.}
}
\rotatebox{90}{
\scalebox{0.9}{
\begin{tabular}{cccccccccccccc}
\hline
$a_v$&$k_{v_2}$&$a_s$&$k_{s_2}$& $a_k$&$k_{k_2}$&$a_0$&W&Cong&$\beta~(12)$&$\beta~(13)$&$r_0$& $a_{c,exc}$&$\sigma$  \\	
\hline
15.2393&1.7168&17.5359&1.404&-&-&-&-&-&-&-&1.218&1.2987&0.660 \\
\hline
15.2876&1.7191&17.8984&1.402&-0.584&-&-&-&-&-&-&1.214&1.3361&0.660 \\
\hline
15.2949&1.7206&17.8647&1.411&-&-&-1.222&-&-&-&-&1.212&1.3530&0.659 \\
\hline
15.2752&1.8799&18.0208&2.785&-1.291&47.24&-&-&-&-&-&1.221&1.2934&0.638 \\
\hline
15.7124&1.8602&19.3863&1.999-&-&-&-&-&y&-&-&1.191&1.0850&0.628 \\
\hline
15.2305&1.8138&16.9776&1.972&-&-&-&21.58&-&-&-&1.233&0.9100&0.608 \\
\hline
15.1186&1.7518&16.8486&1.603&-&-&-&-&-&0.678&-&1.241&1.0501&0.600 \\
\hline
15.1174&1.7910&16.6395&1.803&-&-&-&-&-&-&1.514&1.245&0.9082&0.589 \\
\hline
\end{tabular}
}}
\end{table*}

\end{document}